\documentclass[runningheads]{llncs}
\usepackage[T1]{fontenc}
\usepackage{graphicx}
\usepackage{amsmath}
\usepackage{amsfonts}
\usepackage{makecell}

\usepackage[dvipsnames]{xcolor}
\usepackage{todonotes}

\usepackage{rotating}
\usepackage[pass]{geometry}
\usepackage{float}
\usepackage{lscape}

\usepackage{pgfplots}
\pgfplotsset{compat=1.16}

\usepackage[hidelinks]{hyperref}

\newcommand{\feat}[1]{\ensuremath{\mathbf{#1}}}
\newcommand{\repreq}[1]{=_\feat{#1}}        
\newcommand{\nrepreq}[1]{\neq_\feat{#1}}
\newcommand{\elem}[1]{\ensuremath{\mathrm{#1}}}
\newcommand{\stdfunc}[1]{\ensuremath{S_{\feat{#1}}}}
\newcommand{\gclass}{\mathcal{G}}
\newcommand{\zelem}{\ensuremath{\epsilon}}
\newcommand{\reachstruct}{\ensuremath{\mathcal{S}}}

\newcommand{\chemF}[1]{\ensuremath{\mathrm{#1}}}

\usepackage{booktabs}

\definecolor{graphgreen}{HTML}{71d97e}

\begin{document}
\title{Reconciling Inconsistent Molecular Structures from Biochemical Databases}
%
%
\author{Casper Asbjørn Eriksen\inst{1} \and
Jakob Lykke Andersen\inst{1} \and
Rolf Fagerberg\inst{1}\and
Daniel Merkle\inst{1}}
\authorrunning{C. A. Eriksen et al.}
%
\institute{Department of Mathematics and Computer Science, University of Southern Denmark, Odense, Denmark \\
\email{\{casbjorn,jlandersen,rolf,daniel\}@imada.sdu.dk}}

\maketitle              
\begin{abstract}
    Information on the structure of molecules, retrieved via biochemical databases, plays a pivotal role in various disciplines, such as metabolomics, systems biology, and drug discovery.
    However, no such database can be complete, and the chemical structure for a given compound is not necessarily consistent between databases.
    This paper pre\-sents \textsc{StructRecon}, a novel tool for resolving unique and correct molecular structures from database identifiers.
    \textsc{StructRecon} traverses the cross-links between database entries in different databases to construct what we call an identifier graph, which offers a more complete view of the total information available on a particular compound across all the databases.
    In order to reconcile discrepancies between databases, we first present an extensible model for chemical structure which supports multiple independent levels of detail, allowing standardisation of the structure to be applied iteratively.
    In some cases, our standardisation approach results in multiple structures for a given compound, in which case a random walk-based algorithm is used to select the most likely structure among incompatible alternates.
    We applied \textsc{StructRecon} to the \textit{EColiCore2} model, resolving a unique chemical structure for 85.11\,\% of identifiers.
    \textsc{StructRecon} is open-source and modular, which enables the potential support for more databases in the future.

\end{abstract}

\section{Introduction}
\label{sec:intro}

As the volume of available biochemical information grows, databases have become indispensable resources for researchers, enabling advances in various fields, including metabolomics, systems biology, and drug discovery.
These databases are curated and maintained by different organisations and research groups, each employing their own data collection methods, annotation standards, and quality control procedures.
However, as the collective amount of information stored in the databases expands, so does the amount of errors within databases, and in particular, inconsistencies between them \cite{williams2011,Akhondi2012}.
Discrepancies between biochemical databases pose a significant challenge to researchers performing large-scale analyses, in particular when integrating data from multiple databases \cite{williams2012}. 
In this work, we focus on incompleteness and inconsistencies in the chemical structures within and between database entries,
which can pose a significant problem in applications such as drug discovery, quantitative structure-activity relationship, and atom tracing \cite{young2008,fourches2010}.

The entries in each database may contain quantitative information
about the compounds, structural information on these
compounds, as well as references to related entries in other
databases. A starting observation behind our contribution is that the
cross-database references can be traversed in order to get a fuller
view of the properties of a compound.  In the best case, the
databases complement each other, making up for the incompleteness of
each and allowing the identification a chemical structure of each
compound of interest, even if not all of these are contained in any
single database.

\begin{figure}
    \centering
    \includegraphics[width=.5\textwidth]{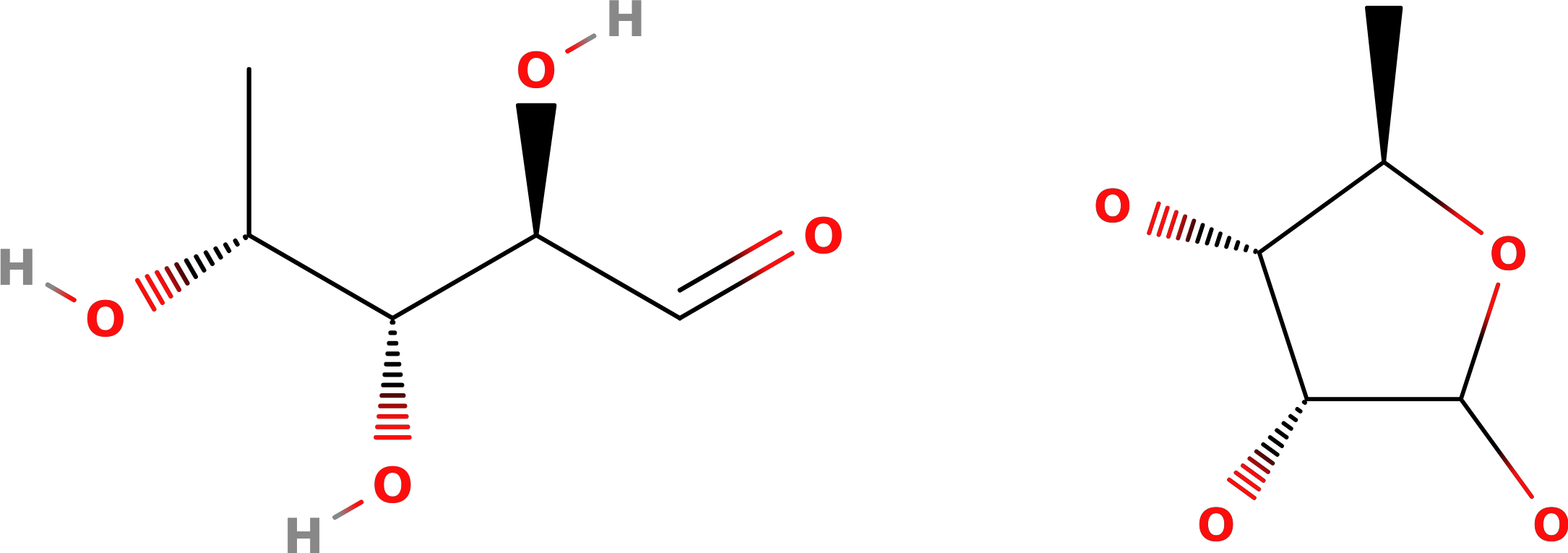}
    \caption{An example of structural discrepancy between databases: the linear (PubChem, ECMDB) and cyclic (ChEBI, MetaNetX, MetaCyc, KEGG) form of 5-deoxy-D-ribose.}
    \label{fig:5ddr}
\end{figure}

However, integrating entries from several databases will invariably introduce discrepancies in the chemical structure. 
In many cases, these discrepancies are simply caused by a difference in the representation of what is intended to be identical chemical structures \cite{muresan_mapping_2012}.
Other times, different structural isomers are present under the same name, for example 5-deoxy-D-ribose, which appears in cyclic and linear forms, depending on the database, as depicted in Fig. \ref{fig:5ddr}.
Over the years, considerable efforts have been directed towards the development of standards and guidelines for chemical structure representation. 
Structural identifiers such as Standard InChI \cite{inchi2015} and Standard SMILES \cite{smiles1,smiles2,smiles3} aim to provide an unambiguous and standardised structural identifier.
However, ambiguity is not completely prevented, as sources may wish to denote chemical structures in varying levels of detail,
e.g., whether to denote stereochemistry, tautomerism, charge, and more. 
SMILES makes no distinction whether such features are explicitly represented, while the layered structure of InChI \cite{inchi-tech-manual} makes it clear which features are to be explicitly represented in some cases.

The problem of comparing chemical structures from entries with different notation and level of standardisation is one of the main challenges of this work.
It is evident that for biochemical problems, an automatic method for retrieving correct chemical structures, up to some degree of standardisation, is needed. 
Our goal is to use the combined resources from several databases in order to create a more complete mapping between database identifiers and chemical structures than any single database provides, while automatically handling discrepancies between these identifiers.

To our knowledge, there are no other tools which give such a consolidated view of the structural information on compounds in databases. 
While some databases, such as ChEBI and PubChem, present data collected from other databases, this is still susceptible to issues such as incompleteness and incorrectness of individual entries.
Furthermore, given a type of identifier, e.g. BiGG, this may not be present in a given database.
For this reason, it is desirable to develop a flexible and extensible system which can in principle be made compatible with any database, while taking into account potential discrepancies.

In Section~\ref{sec:theory}, we establish a \emph{model} for representing the chemical structure of compounds,
with the goal of being able to describe, and compare across, the various levels of detail to which chemical structures are given by databases.
Next, in Section~\ref{sec:method}, we present \textsc{StructRecon}, a \emph{tool} for programmatically retrieving chemical structures from database identifiers,
by traversing database cross-references and using cheminformatics methods for ana\-lysing, comparing, and standardising structural representations based on the model developed in Section~\ref{sec:theory}.
Finally, in section~\ref{sec:results}, we apply the tool to a set of compounds established by genome sequencing of \textit{E. coli} and analyse the resulting network of identifiers and structural representations.

\section{Multi-Level Modelling of Chemical Structures}
\label{sec:theory}
In this section, we introduce a model for chemical structures which allows representation at multiple levels of detail and formalises the standardisation functions which transform structures between these levels.
Compared to established models, such as SMILES and InChI, this model places a particular focus on extensibility, formal specification, and standardisation of structures.

We call the levels of detail \emph{features}.
The seven features used throughout this work will be introduced one-by one as the necessary theory is established.
We categorise identifiers into two classes:
\textit{structural identifiers}, which directly encode a chemical structure, and from which the structure can be recovered algorithmically (e.g.\ InChI, SMILES),
and \textit{symbolic identifiers}, which are generated more or less arbitrarily, and do not carry direct meaning,
but reference an entry in the corresponding database (e.g.\ PubChem CID, BiGG ID).

Depending on the application, the exact definition of a chemical structure may vary.
We wish to model the connectivity of atoms in a molecule, and optionally stereo-chemical information, while the full spatial information is not taken into account.
For this application, the classical method in cheminformatics is the graph-based approach: a structure is, in its most basic form, an unlabelled graph $G = (V,E)$, in which atoms are represented by vertices, and (covalent) bonds are represented by edges.
A graph-based structure is linearly encoded by established structural identifiers, such as SMILES and InChI; even the systematic IUPAC name may be described as a graph-based identifier \cite{IUPAC}.
We may wish to describe additional features, e.g., charge, isotopic labelling, stereochemistry, or tautomerism.
In broad terms, a chemical feature is a class of information about a structure which we may or may not wish to account for when modelling, depending on the input and application. 

We will establish a unified model for graph-based chemical structures.
This provides a consistent view of chemical structure, regardless of the features represented, to which structural identifiers can be mapped.
The underlying simple graph structure is, for our purposes, always assumed to be present, but even basic information such as the chemical element of each atom and the order of covalent bonds, are considered optional features.
First, we formally define the notion of a feature.
\begin{definition}[Feature]
    \label{def:feature}
    A feature $\feat{\Phi}$ is a pair $\feat{\Phi} = (\feat{\Phi}_V, \feat{\Phi}_E)$,
    where $\feat{\Phi}_V$ and $\feat{\Phi}_E$ are sets of possible values for the attribute on atoms and bonds respectively, each of which must contain a special `nil'-element, $\zelem$, indicating that the value is unspecified or not applicable. 
\end{definition}
We will define and apply seven such features in this work. Starting with the most essential, the element of each atom can be expressed as a feature \feat{E}, with
$$\feat{E} := (\{\zelem, \elem{H}, \elem{He}, \elem{Li}, \dots \}, \{\zelem\}). $$ 
For a given structure, each vertex will either be assigned \zelem, indicating that no indication is given as to the element of that atom,
or it will be assigned a specific element.
Edges can only be assigned \zelem, indicating that this feature assigns no attribute to edges.
Similarly, bond types are expressed as the feature
$$\feat{B} := (\{\zelem\}, \{ \zelem, -, =, \equiv \}).$$
This feature can be expanded to also indicate other bond types, such as aromatic or ionic, if needed.
The isotope of each atom can be stated as
$$\feat{I} := (\{\zelem\} \cup \mathbb{N}, \{\zelem\}),$$
where the vertex attribute indicates the atomic weight of each atom.
Before describing the remaining features, we need a precise definition of chemical structure. Combining the sets of values for all features, we can define the overall feature space, which is needed for a formal definition of a chemical structure.
\begin{definition}[Feature space]
    \label{def:featurespace}
    Given a set of features $\{\feat{\Phi}_1, \dots, \feat{\Phi}_n\}$, the feature space for vertices and edges,
    $\mathcal{F}_V$ and $\mathcal{F}_e$, respectively, is the combined attribute space of the features,
    where $\mathcal{F}_V = \feat{\Phi}_{1V} \times \dots \times \feat{\Phi}_{nV}$ and $\mathcal{F}_E = \feat{\Phi}_{1E} \times \dots \times \feat{\Phi}_{nE}$.
\end{definition}
\begin{definition}[Chemical structure]
    \label{def:chemstructure}
    A chemical structure is an undirected graph $G = (V, E, A_V, A_E)$, where $V$ is the set of atoms, $E$ is the set of covalent bonds,
    and $A_V\colon V \to \mathcal{F}_V$ (resp.\ $A_E\colon E \to \mathcal{F}_E$) is an attribute function, assigning to each vertex (resp.\ edge) a value for each feature in the  feature space.
\end{definition}
Let $\mathcal{G}$ be the set of all such chemical structures.
Simultaneously working in several levels of detail, i.e., features,
naturally raises the problem of how to compare equivalence of structures between different sets of features. 
For this, we introduce, for each feature, a \emph{standardisation function}.
\begin{definition}[Standardisation function]
    \label{def:stdfunction}
    Given a feature \feat{\Phi}, the standardisation function w.r.t.\ \feat{\Phi} is $\stdfunc{\Phi}\colon \gclass \to \gclass$.
    The function is required to be idempotent, and all vertices and edges of the resulting structure should have $\zelem$ as value for feature \feat{\Phi} in the attribute function.
\end{definition}
We extend the definition to sequences of features, resulting in a composition of standardisations, i.e.,
$\stdfunc{\Phi_1 \dots \Phi_n} = \stdfunc{\Phi_n} \circ \dots \circ \stdfunc{\Phi_1}$.

For a structure $G$, the image $\stdfunc{\Phi}(G)$ is the corresponding standardised structure, which does not contain any information about feature \feat{\Phi}.
For a sequence of features $\feat{\Phi_1}, \dots, \feat{\Phi_n}$, we define $\mathcal{G}_{\feat{\Phi_1} \dots \feat{\Phi_n}} \subseteq \mathcal{G}$
as the set of structures which are standardised according to the of features.
That is, all applicable atom and bond attributes are $\zelem$ these features, and they are their own image in the standardisation function: 
$$G \in \mathcal{G}_{\feat{\Phi_1} \dots \feat{\Phi_n}} \iff \stdfunc{\Phi_1 \dots \Phi_n} \- (G) = G.$$ 

For the features described so far, \feat{E}, \feat{B}, and \feat{I}, the \emph{trivial standardisation function} is sufficient.
This function simply erases the attributes by setting them to $\zelem$.
In some cases, the trivial standardisation function is not sufficient. For example, we define the charge feature as
$$\feat{C} := (\{\zelem\} \cup \mathbb{Z}, \{\zelem\}),$$
where the vertex attribute indicates the charge of the atom.
The process of standardising the charge of a given molecule may be limited by chemical constraints, such as is the case in RDKit~\cite{rdkit} and InChI~\cite{inchi-tech-manual}.
The function may, among other modifications, add and/or remove hydrogen atoms, trying to remove charges from each atom in a chemically valid way.
We will not further discuss the intricacies of this operation, as it is implementation-dependant.~

For some features, the definition even depends entirely on the standardisation function. E.g., the fragment feature~\cite{muresan_mapping_2012}:
$$\feat{F} := (\{\zelem\}, \{\zelem\}).$$
In this case, no information is explicitly encoded in the graph, but we still want a way to standardise w.r.t. fragments.
We define $\stdfunc{F}(G)$ to be the largest connected component of $G$ (measured by the number of non-hydrogen atoms), with an implementation-specific method for breaking ties based on other attributes of each connected component.

Tautomerism is similarly difficult to define due to the complexities involved in determining the tautomers of any given structure. 
Let 
$$\feat{T} := (\{\zelem\}, \{\zelem\}).$$
Again, we have no features, instead relying on the standardisation function: assume a chemical oracle,
which given a structure $G$, returns the set of all tautomeric structures, according to some definition. 
Then, let a deterministic method choose a canonical representative from among these structures.
The standardised structure $\stdfunc{T}(G)$ is this canonical tautomer.

For stereochemistry, there are multiple methods for encoding information about local geometry at the vertex and edge level~\cite{cheminfstereo,stereo2,stereo3}
Any method for encoding such information can be used to generate the feature \feat{S}. 
The trivial standardisation function will in most cases be sufficient for standardising structures with respect to stereochemistry.

Finally, we define equivalence of chemical structures:
With the notation $G_1 \repreq{\feat{\Phi_1} \dots \feat{\Phi_n}} G_2$, we denote that the structures $G_1$ and $G_2$ are equivalent up to standardisation of $\feat{\Phi_1}, \dots, \feat{\Phi_n}$, 
i.e., they are equal when the features $\feat{\Phi_1}, \dots, \feat{\Phi_n}$ are not considered.
Formally 
$$G_1 \repreq{\Phi_1 \dots \Phi_n} G_2 \iff \stdfunc{\Phi_1, \dots, \Phi_n} \, (G_1) = \stdfunc{\Phi_1, \dots, \Phi_n} \, (G_2) .$$
As an example, consider the structures for methanol and methoxide,
where $\chemF{CH_3OH} \nrepreq{FICTS} \chemF{CH_3O^-}$, but $\chemF{CH_3OH} \repreq{FITS} \chemF{CH_3O}^-$.

We have now described the features which are considered in this contribution, based on the FICTS features \cite{sitzmann2008}: 
(\feat{E}) elements, (\feat{B}) bonds, (\feat{F}) fragments, (\feat{I}) isotopes, (\feat{C}) charge, (\feat{T}) tautomerism, and (\feat{S}) stereoisomerism.
It should be noted that some of these features depend upon each other, e.g.,
it would not make sense to specify the isotope of an atom without also specifying the element.
The standardisation functions can not be expected commute with each other in general.
For this reason, \textsc{StructRecon} needs a defined order in which the standardisation functions will be applied.

\section{Algorithms and Implementation}
\label{sec:method}
In this section, we describe the ideas and algorithms of \textsc{StructRecon}, based on the model developed in Section~\ref{sec:theory}.

\subsubsection{Data sources}
We used six sources of data: BiGG~\cite{BiGG}, ChEBI~\cite{chebi}, the E.\ Coli Metabolome Database (ECMDB)~\cite{ECMDB1,ECMDB2}, KEGG~\cite{kegg}, MetaNetX~\cite{metanetx}, and PubChem~\cite{pubchem2023}.
These were selected based on their programmatic accessibility and relevance to metabolic modelling.
\textsc{StructRecon} is modular, making it easy to add more data sources in the future.


MetaNetX uses a versioning system in which entries present in one version are not necessarily present in another. 
Inter-database references do not typically specify which version of MetaNetX is referenced, but MetaNetX keeps a record of deprecated entries. 
Following the chain of deprecations, it is possible to obtain the newest entries corresponding to any given entry. 
The deprecation relationship can both split and merge entries in between versions. 

\subsubsection{Construction of the Identifier Graph}
The interconnected nature of identifiers within and between databases is represented as a directed graph, called the \textit{identifier graph}. 
In the identifier graph, vertices correspond to identifiers, while arcs represent relationships between these.
Each vertex contains as attributes, the type of ID (e.g.\ PubChem CID, BiGG ID, SMILES) called the \emph{identifier class}, as well as the actual ID.
Each edge is annotated with the source database.

A number of \emph{procedures} are specified, each being a subroutine which takes identifiers as input, and finds associated identifiers in chemical databases.
By executing the procedures in an order which seeks to minimise overhead, the identifier graph is built iteratively, starting with the \emph{input vertices} which are directly obtained as input. 
The resulting graph will contain symbolic identifiers as well as the structural identifiers as they appear in the respective databases. 
For an example of a complete identifier graph, refer to Fig. \ref{fig:reach_5mtr}.

\subsubsection{Structure Standardisation}
When a structural identifier (SMILES, InChI) is added to the identifier graph, it is first converted to an internal graph-based representation,
in accordance with Def.~\ref{def:chemstructure}.
From this point, we assume that the atom (\feat{A}) and bond (\feat{B}) features are always implicitly represented, and will therefore refer to the remaining features as \feat{FICTS}. 
As the standardisation functions are not expected to commute, in general, we enforce a particular order on the features defined by the user, by default $\feat{F}, \feat{I}, \feat{C}, \feat{T}, \feat{S}$.
This was chosen as the default ordering, as it produces the greatest number of uniquely resolved structures which a lesser degree of standardisation.

We assign to each structural identifier $G$, an attribute specifying for which features a structure is standardised. 
For features \feat{F}, \feat{I}, and \feat{S}, it is simple to guarantee that it is standardised by inspecting the structure.
Checking \feat{F} is simply examining the connectivity of the graph, and $\feat{I}$, \feat{S}, inspecting for all vertices and edges whether they have the equivalent of $\zelem$ as attribute.
For \feat{C} and \feat{T}, we check whether a structure is standardised by applying the respective standardisation functions.

When the links to new databases is exhausted, in many cases, there will be multiple different structural identifiers associated with each compound.
We aim to achieve a unified representation of the compounds by iteratively applying the standardisation functions:
For any structure $G$, which is not fully standardised, that is, $G \notin \gclass_{\feat{FICTS}}$, let $\feat{\Phi_k}$ be the first feature in which $G$ is not standardised according to the feature ordering. 
That is, $\stdfunc{\feat{\Phi_1} \dots \feat{\Phi_{k-1}}}(G) = G$, but $\stdfunc{\feat{\Phi_1} \dots \feat{\Phi_{k}}}(G) \not= G$. Then produce $G' = \stdfunc{\Phi_k}(G)$, adding this new structure to the identifier graph, with an arc $(G, G')$. If $G'$ is equivalent to an existing structure $H$, then no new vertices are created, but the arc $(G, H)$ is added instead.
The standardisation process is visualised in the output of \textsc{StructRecon} (Fig. \ref{fig:reach_5mtr}) in which the blue nodes represent structures, and purple nodes represent maximally standardised structures.

\subsubsection{Structure Selection}
While the identifier graph is a general digraph, considering only the structural identifiers yields a forest of in-trees, as standardisation functions are many-to-one, and may therefore merge structures, but never split. 
For each input, the transitive and reflexive closure of the forest of in-trees imposes a partial order on the reachable structures. 
In this partial order, structure $G_1$ precedes $G_2$ if $G_2$ is a more standardised identifier, reachable from $G_1$ by applying standardisation functions.
Maximal elements are fully standardised structures, $G \in \gclass_\feat{FICTS}$.

For each input identifier $i$, \textsc{StructRecon} should resolve the input to a single structure.
The vertices reachable from $i$  represent symbolic and structural identifiers which are related to $i$ through database links, as well as structural identifiers which can be derived from these by standardisation. Denote by $\reachstruct(i)$ the partial order of structures reachable from $i$.

If $\reachstruct(i)$ has one maximal (greatest) element, then we say that $\reachstruct(i)$ is resolved and \emph{consistent} --- all sources for the compound can agree on a structure, at least up to the highest degree of standardisation.
That is, $G \repreq{FICTS} G'$ for any pair of structures $G$, $G'$ in the reachable database entries.
If there are multiple maximal elements, then the sources cannot agree, and we call $\reachstruct(i)$ \emph{inconsistent}.
If $\reachstruct(i)$ contains no structures at all, then the compound is \emph{unresolved}.
In the case where $\reachstruct(i)$ is consistent, we want to select the most specific element on which all sources agree. 
In the partial order, this is the supremum of the set of structures which were found directly in the databases. 

If $\reachstruct(i)$ is inconsistent, resolving $i$ requires a choice between the maximal structures. 
This choice can be made automatically by a scoring algorithm which, for each input, assigns a confidence score to all vertices reachable from the input vertex.
The scoring algorithm essentially computes the probability that a random walk in the identifier graph, starting at the input vertex, will arrive at each vertex. 
The algorithm is based on PageRank~\cite{pagerank}, with the key differences that the initial probability distribution is $1$ for the input vertex, and $0$ for all other vertices, and that sink vertices only loop back to the input vertex, rather than all vertices. 
After assigning a confidence score to each vertex, the confidence scores of each maximally standardised structure is evaluated.
The \emph{confidence ratio} is computed, the confidence of the second-most likely structure over the confidence of the most likely structure.
If this value is below a given threshold (0.5 by default), then the vertex with the highest confidence score is automatically selected. Otherwise, no structure can be chosen with high enough confidence, and the input is marked for manual disambiguation.

\subsubsection{Implementation}
\textsc{StructRecon} was implemented in Python, and is available at \url{https://github.com/casbjorn/structrecon}. 
The accompanying web interface can be accessed at \url{https://cheminf.imada.sdu.dk/structrecon}.
Our model for chemical structure is implemented using RDKit~\cite{rdkit}, which furthermore provides functions for uncharging structures and computing the canonical tautomer.

%
%
%
%
\section{Results}
\label{sec:results}
The tool was tested on the metabolic network model \textit{EColiCore2} \cite{ECC2}.
The model contains 2138 identified compounds, with associated BiGG IDs
We chose this dataset for evaluation, as the selected databases have a particular focus on biochemistry,
and because of the well-established nature of the \textit{E.\ coli} genome.

Of the 2138 inputs, 136 (6.36\,\%) were identified as macromolecules,
based on string-matching BiGG IDs and names found in databases to an incomplete list of substrings associated with macromolecules,
such as ``tRNA'' and the names of various proteins and enzymes.
We consider the handling of macromolecules to be out of the scope of this work.
In the identifier graph, an average of 31.70 vertices are reachable from each input vertex.
Of the non-macromolecule inputs, 1459 (72.88\,\%) resolved to exactly one structure up to maximal \feat{FICTS} standardisation, while only 492 (24.58\,\%) had only one structure up to \feat{FICT} standardisation.
\begin{figure}[tb!]
    \centering
    \includegraphics[width=\textwidth]{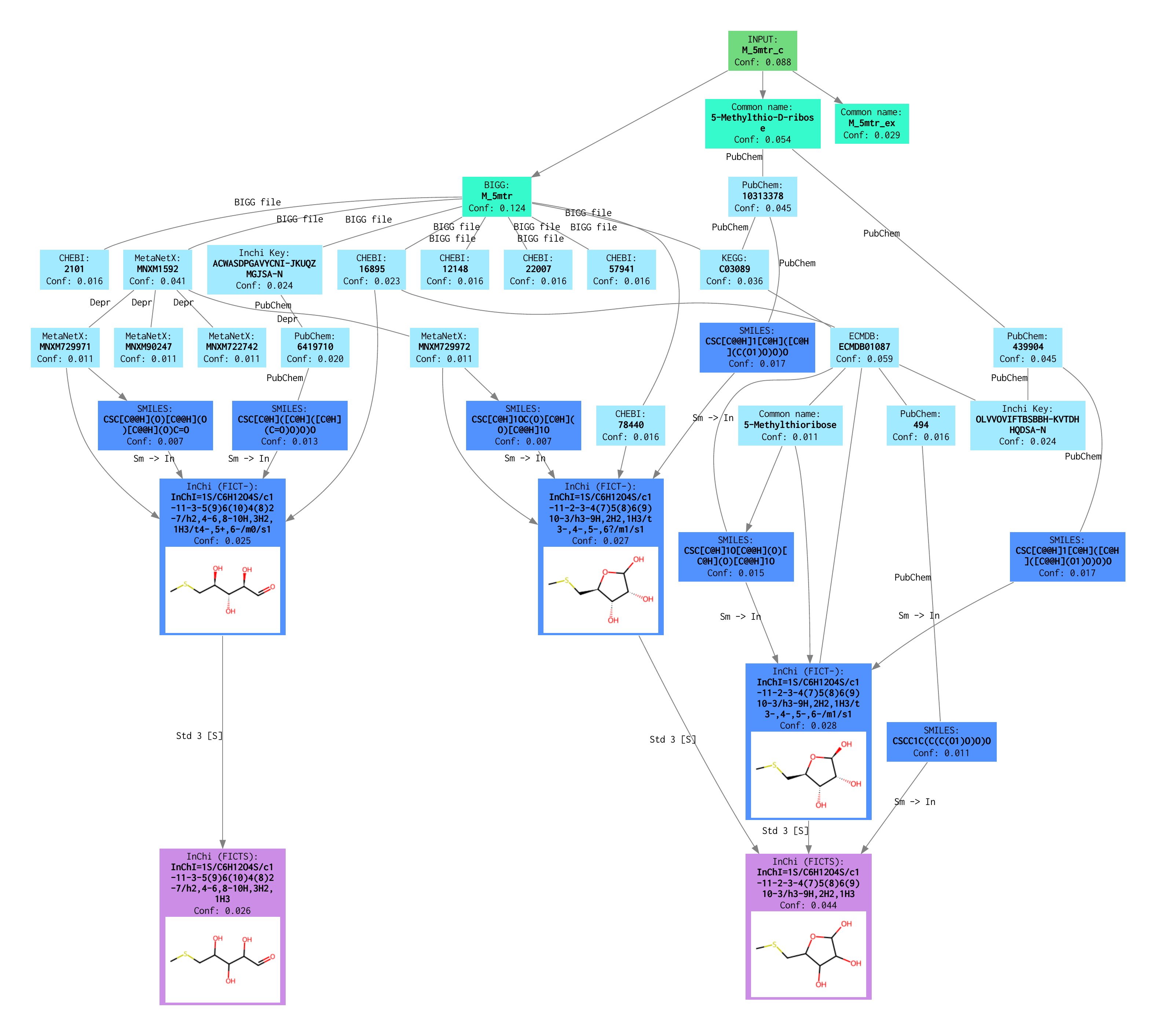}
    \caption{
        The identifier graph generated by the BiGG ID \texttt{M\_5mtr\_c}.
        Each vertex displays the type of identifier, the identifier itself, and the confidence assigned by the scoring algorithm. 
        For structures, the set of features in which the structure is standardised is also displayed, along with a graphical representation.
        The green vertex is the input vertex.
        The turquoise vertices are symbolic identifiers found directly within the input file, in this case the \textit{EColiCore2} model.
        The light blue vertices are other symbolic identifiers.
        The dark blue vertices represent structural identifiers, either found in databases, or obtained by standardisation.
        The violet vertices represent maximally standardised structures.
        Arcs with no direction are shorthand for one arc in each direction.
    }
    \label{fig:reach_5mtr}
\end{figure}
Of the non-macromolecule inputs, 57 (2.85\,\%) yielded no structure at all.
Examples of this category includes 
bis-molybdopteringuaninedinucleotide (BiGG: \verb!M_bmocogdp!), 
Hexadecanoyl-phosphate(n-C16:1) (BiGG: \verb!M_hdceap!),
and 2-tetradec-7-enoyl-sn-glycerol-3-phosphate (BiGG: \verb!M_2tdec7eg3p!).

\begin{figure}[tb!]
    \centering
    \begin{tikzpicture}
        \begin{axis}[
            width=0.85\textwidth,
            height=4cm,
            ymin=0, ymax=500,
            ytick distance = 100,
            minor y tick num = 0,
            ytick pos = left,
            ylabel style={align=center},
            ylabel = Unique\\resolution count,
            ymajorgrids = true,
            yminorgrids = true,
            ytick style={draw=none},
            xtick style={draw=none},
            xmin = 0.05, xmax = 1.05,
            xtick distance = 0.1,
            xlabel = Confidence ratio,
            axis x line = bottom,
            x axis line style={opacity=0},
            y axis line style={opacity=0},
            nodes near coords,
            nodes near coords style={black}
            ]
            \addplot+[
                graphgreen,
                fill = graphgreen,
                ybar,
                mark=no
                ] plot coordinates { 
                (0.1, 8) (0.2, 29) (0.3, 91) (0.4, 185 ) (0.5, 245) (0.6, 305)
                (0.7, 334) (0.8, 369) (0.9, 427) (1.0, 486)
                };
        \end{axis}
    \end{tikzpicture}
    \caption{
        For several choices of confidence ratio threshold (the confidence of the second-most likely structure over the confidence of the most likely structure), shows the number of inputs, out of 486, which resolve to a unique structure in the ECC2 model. \textsc{StructRecon} uses a default threshold of 0.5. Setting the threshold to 0.0 would mean only choosing a structure if no alternatives exist, while a threshold of 1.0 results in picking one of the structures with the highest confidence arbitrarily.
    }
    \label{fig:conf-threshold-histogram}
\end{figure}
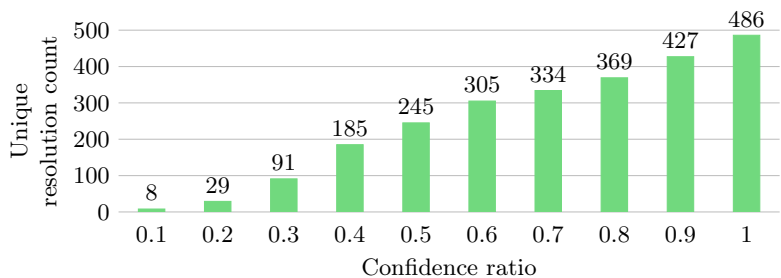

A total of 486 inputs (24.28\,\%) yielded multiple maximally standardised structures, and needed to be disambiguated based on the \emph{confidence ratio}.
In our experimentation, we found 0.5 to be a reasonable threshold, meaning that we select the structure with the highest confidence if it has at least twice the confidence of any other structure.
With a threshold of 0.5, an additional 245 inputs were uniquely resolved, for a total of 1704 consistent inputs (85.11\,\%),
leaving 241 compounds for manual disambiguation. The effect of different choices of confidence ratio threshold is displayed in Fig.~\ref{fig:conf-threshold-histogram}.

We will proceed to describe some concrete examples of identifier graphs which serve to demonstrate both the problem
of database inconsistency and the solution provided by \textsc{StructRecon}.
One example is 5-methylthio-D-ribose. 
The associated identifier graph is displayed in Fig.~\ref{fig:reach_5mtr}.
Database interconnections do not necessarily make distinctions between this compound and S-methyl-5-thio-D-ribose. 
The confidence ratio between these two maximised structures is 0.59,
indicating a relatively high degree of interconnection between the associated database entries. 
While the correct structure has the highest confidence score, the default threshold of 0.5 would mark this discrepancy for manual disambiguation.

Unexpectedly, the simplest and most prevalent molecules turns out to be inconsistent, but easy to reconcile based on the confidence ratio. 
A good example is water, as displayed in Appendix~A. 
The conventional structure \verb|H2O| is found in a multitude of databases, however, the ChEBI identifier \verb|29356| (oxide(2-)) is associated with the generic BiGG identifier for water through the BiGG database. 
However, as this is the only connection, that structure is assigned a smaller confidence score than the conventional structure by the scoring algorithm. 
This graph therefore has a low confidence ratio of 0.07, representing a high degree of support of the conventional structure, which is chosen by \textsc{StructRecon}.

\section{Conclusion}
\label{sec:conclusion}
In this work, we propose a model for chemical structure, which supports multiple levels of standardisation.
Based on this model, we present \textsc{StructRecon}, a novel tool which identifies and reconciles the chemical structure of compounds
based on the traversal of interconnections between biochemical databases.
We applied the tool to \textit{EColiCore2}, a metabolic model of \textit{E.\ coli}.
In 85.11\,\% of cases, a chemical structure could be uniquely identified with reasonable confidence,
demonstrating that \textsc{StructRecon} can be a valuable tool for structure-based approaches in bioinformatics and related fields.
\textsc{StructRecon} is open-source and developed with modularity in mind, making integration of additional databases and procedures possible.

\subsubsection*{Acknowledgements} This work is supported by the Novo Nordisk Foundation grants NNF19OC0057834 and NNF21OC0066551.

%
%
%
\bibliographystyle{splncs04}
\bibliography{refs}

\appendix
\newgeometry{left=2cm,right=2cm,top=2cm,bottom=2cm}
\thispagestyle{empty}

\begin{landscape}

\section*{Appendix A}
\label{app:water}
\begin{figure}[h!]
    \centering
    \includegraphics[width=24cm]{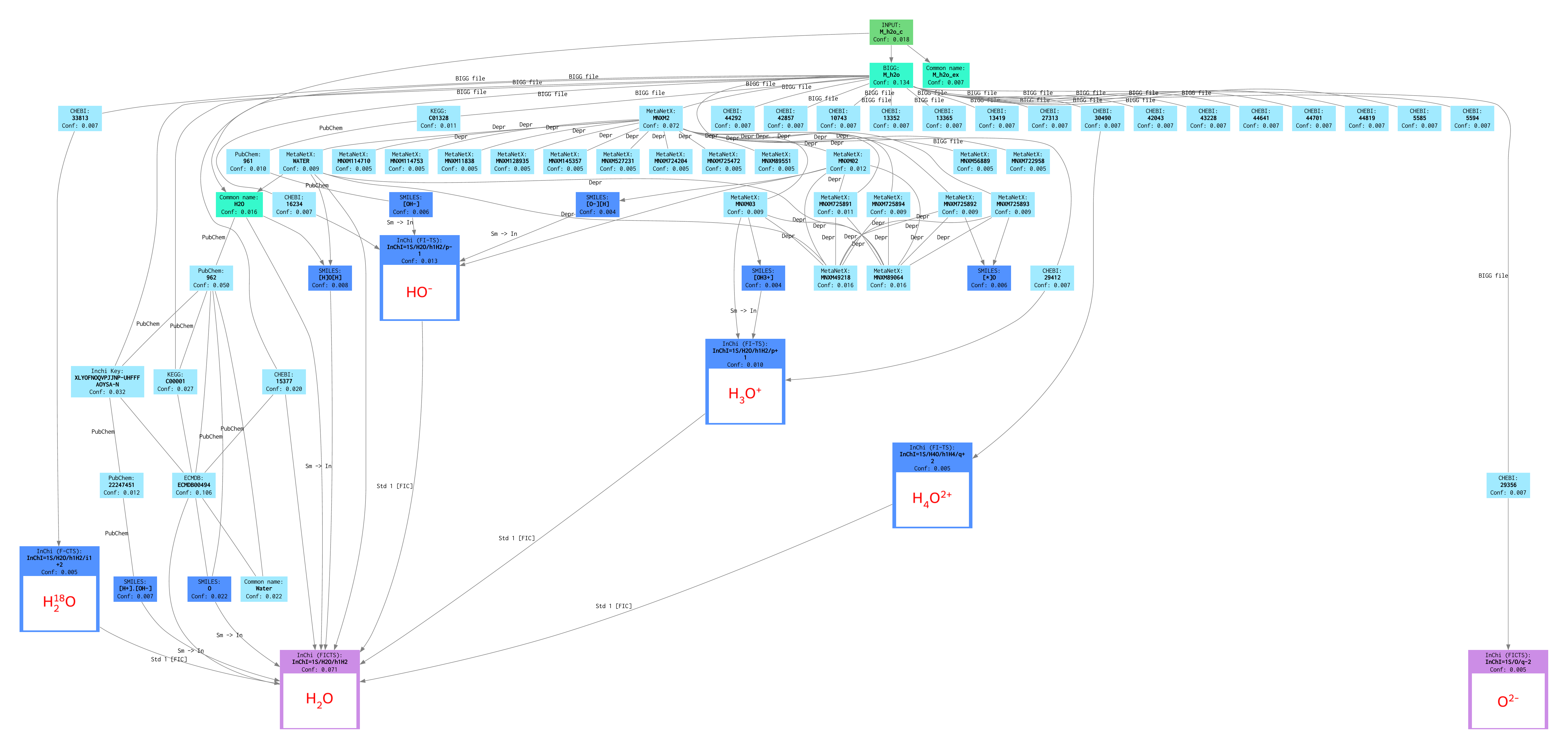}
    \caption{
        The identifier graph generated by the BiGG ID \texttt{M\_h20\_c}.
        For a description of how to read this graph, see the caption of Fig. \ref{fig:reach_5mtr}.
        This figure clearly demonstrates the impact of the scoring algorithm, as it chooses the conventional structure, \chemF{H_2O}, at a confidence ratio of 0.07.
    }
    \label{fig:reach_water}
\end{figure}

\end{landscape}
\restoregeometry

\end{document}